# Seismic Earth Pressure Development in Sheet Pile Retaining Walls: A Numerical Study

P. Rajeev[1*], H. Bui[2], N. Sivakugan[3]

[1] *Department of Civil and Construction Engineering, Swinburne University of Technology, Australia*
[2] *Department of Civil Engineering, Monash University, Australia*
[3] *School of Engineering & Physical Sciences, James Cook University, Townsville, Qld, Australia*

*Corresponding author. Email: prajeev@swin.edu.au*

**Abstract:** The design of retaining walls requires the complete knowledge of the earth pressure distribution behind the wall. Due to the complex soil-structure effect, the estimation of earth pressure is not an easy task; even in the static case. The problem becomes even more complex for the dynamic (i.e., seismic) analysis and design of retaining walls. Several earth pressure models have been developed over the years to integrate the dynamic earth pressure with the static earth pressure and to improve the design of retaining wall in seismic regions. Among all the models, Mononobe-Okabe (M-O) method is commonly used to estimate the magnitude of seismic earth pressures in retaining walls and is adopted in design practices around the world (e.g., EuroCode and Australian Standards). However, the M-O method has several drawbacks and does not provide reliable estimate of the earth pressure in many instances. This study investigates the accuracy of the M-O method to predict the dynamic earth pressure in sheet pile wall. A 2D plane strain finite element model of the wall-soil system was developed in DIANA. The backfill soil was modelled with Mohr-Coulomb failure criterion while the wall was assumed behave elastically. The numerically predicted dynamic earth pressure was compared with the M-O model prediction. Further, the point of application of total dynamic force was determined and compared with the static case. Finally, the applicability of M-O methods to compute the seismic earth pressure was discussed.

## 1. Introduction

Regardless the multitude of studies that have been carried out over the years, the dynamic response of earth-retaining walls is far from being well understood. There is, in current engineering practice, a lack of conclusive information that can be used in design method. The most commonly used methods to design earth-retaining structures under seismic conditions are force-based equilibrium approaches like the pseudo-static analysis (e.g. Mononobe-Okabe [1]) and pseudo-dynamic techniques (Steedman and Zeng [2]), and displacement-based procedures such as the sliding block method (e.g. Richards and Elms [3]). In the limit-state methods of analyses, the wall is considered to displace or deform sufficiently at the base to fully mobilize the shearing strength of the backfill.

Even under static conditions, prediction of actual retaining wall pressures and deformations constitute a complicated soil-structure interaction problem. The dynamic response of even the simplest type of retaining wall is therefore a quite complex phenomenon. It depends on the mass and stiffness of the wall, the backfill and the underlying ground, as well as the interaction among these components and the nature of the seismic motions.

The purpose of this study was to develop a finite element model to shed light into understanding the dynamic behaviour of sheet pile wall, in particular to find the distribution of dynamic lateral earth pressures. In all the analyses, the soil was assumed to behave as a homogeneous, elasto-plastic medium with a Mohr-Coulomb failure criterion. The wall was assumed to behave as a linear elastic material. The numerical model for the wall and surrounding soil have been developed using DIANA [4], a commercially available finite element program.

The results obtained with DIANA were compared with results obtained from pseudo-static analysis using the procedure by Mononobe-Okabe and Wood analytical solutions.

## 2. Mononobe-Okabe Earth Pressure Models

Okabe [5], Mononobe and Matsuo [1] were the pioneers to obtain the active and passive earth pressure coefficients under seismic conditions. It was an extension of Coulomb's method in the static case for determining the earth pressures by considering the equilibrium of a triangular failure wedge. The method is now commonly known as Mononobe-Okabe method. For active and passive cases, planar rupture surfaces were assumed in the analysis. Figure (1) shows the failure surfaces at active and passive states, and the forces considered in the analysis.

The Mononobe-Okabe approach is valuable in providing a good assessment of the magnitude of the peak dynamic force acting on a retaining wall. However, the method is based on three fundamental assumptions:

1. The wall has already deformed outwards sufficiently to generate the minimum (active) earth pressure;
2. A soil wedge, with a planar sliding surface running through the base of the wall, is on the point of failure with a maximum shear strength mobilized along the length the surface; and
3. The soil behind the wall behaves as a rigid body so that acceleration can be assumed to be uniform throughout the backfill at the instant of failure.

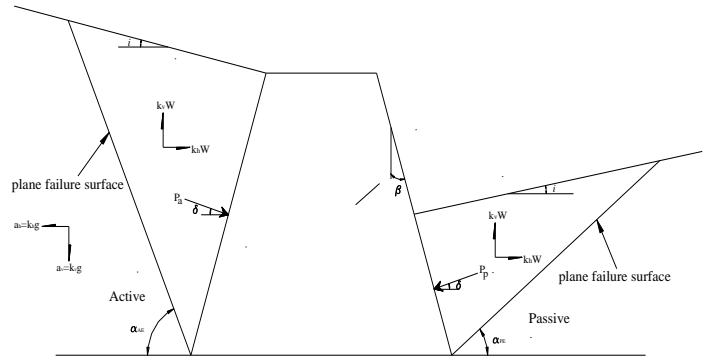

**Figure 1:** Failure surfaces and the forces considered by Mononobe-Okabe

The expression for computing the seismic active and passive earth force, $P_{ae,pe}$, is given by

$$P_{ae,pe} = \frac{1}{2}\gamma H^2 (1-k_v) K_{ae,pe} \qquad (1)$$

$$K_{ae,pe} = \frac{\cos^2(\phi \mp \beta - \theta)}{\cos\theta \cos^2\beta \cos(\delta \pm \beta + \theta)\left[1 \pm \left(\frac{\sin(\phi+\delta)\sin(\phi \mp i - \theta)}{\cos(\delta \pm \beta + \theta)\cos(i-\beta)}\right)^{0.5}\right]^2} \qquad (2)$$

where, $\gamma$ is the unit weight of soil, $H$ is the vertical height of the wall, $K_{ae,pe}$ is the seismic active and passive earth pressure coefficient, $\varphi$ is the soil friction angle, $\delta$ is the wall friction angle, $\beta$ is the wall inclination with respect to vertical, $i$ is the ground inclination with respect to horizontal on both sides of the wall, $k_h$ is the seismic acceleration coefficient in the horizontal direction, and $k_v$ is the seismic acceleration coefficient in the vertical direction.

$$\theta = \tan^{-1}\left[\frac{k_h}{1-k_v}\right] \qquad (3)$$

In Mononobe-Okabe analysis, the point of application of the total seismic active or passive force is considered to be at $H/3$ from base of the wall, but experimental results (Jacobse 1939, Matsuo 1941)

[6] show it is slightly above *H*/3 from base of the wall for seismic active case. Prakash and Basavanna [7] have made an analysis to determine the height of the resultant force in the Mononobe-Okabe analysis. Seed and Whitman [8] recommended that the dynamic component be taken as acting at 0.6*H*. Mononobe-Okabe analyses show that $k_v$, when taken as one-half to two-thirds the value of $k_h$, affects the total active or passive pressure by less than 10%. Seed and Whitman [8] concluded that vertical accelerations can be ignored when the Mononobe-Okabe method is used to estimate the total pressure for typical wall designs.

The Mononobe-Okabe method is very simple and straightforward, has been used by designers for long, because experimental and theoretical studies have shown that it gives satisfactory results in cases where the backfill deforms plastically and the wall movement is large and irreversible (Whitman [9]). However, there are many practical cases, such as massive gravity walls or basement walls braced at top and bottom, where the wall movement is not sufficient to induce a limit state in the soil.

## 3. Sheet Pile Wall-Soil System

Figure 2 shows the soil-wall system that has been studied in this paper. The height of the flexible wall is 6.0 m, with 5 m of embedment. The backfill and foundation soil is assumed to be medium-dense, cohesion-less, compacted fill. Its geotechnical properties are as follows: unit weight: $γ_s$ = 19.6 kN/m³; effective angle of internal friction: $φ'$ = 40°. The water table is assumed located well below the bottom of the wall and thus the analyses are performed assuming dry soil. It is a plane strain problem.

The properties of the concrete and of the reinforcing steel used for designing the wall are as follows: concrete unit weight : $γ_c$ = 23.6 kN/m³; concrete compressive strength: $f'_c$ = 27.6 MPa; steel yield strength: $f'_y$ = 413.4 MPa.

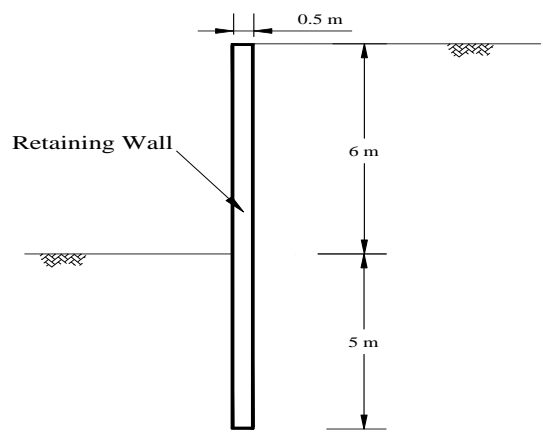

**Figure 2: Dimensions of the sheet pile retaining wall**

The primary parameters governing the dynamic response of the system are the relative flexibility of the wall and retained medium and relative flexibility of the rotational point constrain given by retained soil. The characteristics of the base motion also affect the response.

## 4. Numerical Model

The DIANA finite element model consists of the upper 15 m of the wall-soil system; containing wall and backfill and 4 m of the underlying natural soil below the base of the wall. Laterally, the model is approximately 38.5 m, to include 12 m of existing soil in front of the wall, approximately 26 m of the backfill/existing soil behind the wall and 0.5 m wall thickness (Figure 3).

The soil and wall are modelled using eight-node quadrilateral isoparametric plane strain elements. These elements are based on quadratic interpolation of displacement and Gauss integration. An elasto-plastic constitutive model, in conjunction with Mohr-Coulomb failure criterion, is used to model the soil. Plane-strain elements are also used to model the concrete retaining wall as a linear elastic material. The wall/backfill was "numerically constructed" in DIANA similar to the way an actual wall would be constructed. The soil in front of the wall is excavated in two steps, each of 3 m, with the model being brought to static equilibrium after each excavation. Such excavation allowed realistic earth pressure to develop as the wall deformed during the excavation.

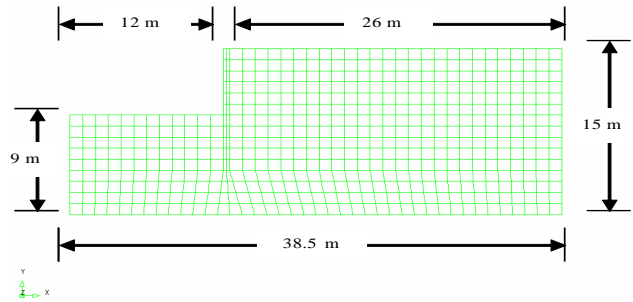

**Figure 3. Annotated DIANA model of the wall-soil system**

The small strain natural frequency of the DIANA model of the retaining wall-soil system is estimated to be 4.8 Hz (≈ 5.0 Hz). At higher stains, it is expected that the natural frequency of the system will be less than 5 Hz. The cut-off frequency for dynamic analysis was set at 15 Hz. All along the model, the size of the elements varied from 0.25 to 1.0 m in both directions, which was less than one eighth of the shortest wave length that corresponds to the highest frequency of 15 Hz considered in the transient analysis (Kuhlemeyer and Lysmer [10]). Total of 806 elements was used in the model.

### 4.1. Model Parameters for the Soil

The stress-strain behaviour of the soil was modelled using the Mohr-Coulomb constitutive model. The soil input parameters used in the finite element model are given in Table 1.

**Table 1. DIANA input properties of sand**

| Parameters | value |
|---|---|
| Poisson's ratio | 0.26 |
| At-rest pressure coefficient | 0.36 |
| Small strain Young's modulus (MPa) | 163.13 |
| Effective friction angle | 40º |
| Density (kg/m$^3$) | 2000 |

### 4.2. Model Parameters for the Wall

The concrete wall is modelled to behave linear elastically during the whole analysis. The parameters are required to define the mechanical properties of the wall: density ($\rho$), elastic modulus ($E_c$), and Poisson's ratio ($v$). $E_c$ is calculated using following equation:

$$E_c = 5000\sqrt{f_c^{'}} \quad \text{MPa} \qquad (4)$$

where $f'_c$ is compressive strength of concrete.

**Table 2. DIANA input properties of concrete**

| Parameters | Value |
|---|---|
| Elastic modulus of concrete (MPa) | 30,000 |
| Yield strength of steel (MPa) | 413.400 |
| Young's modulus of steel (GPa) | 200 |
| Density of concrete (kg/m$^3$) | 2400 |

### 4.3. Damping

As stated before, the soil was modelled as an elasto-plastic Mohr-Coulomb material. Inherent in this model is that once the induced dynamic shear stresses exceed the shear strength of the soil, the plastic deformation of the soil introduced considerable hysteretic damping. However, for dynamic stresses less than the shear strength, the soil behaves elastically, without any damping. In order to avoid over damping in large deformation situations, a lower bound damping ratio of one percent of

Rayleigh damping was set to the soil at first natural frequency of the system and the predominant frequency of the excitation.

## 5. Ground Motion

To perform the dynamic analysis, three real earthquake acceleration time-histories were selected with varying levels of intensity and they include the 1940 Imperial Valley earthquake (California), the 1999 Chi-Chi earthquake (Taiwan), and the 1995 Hyogoken-Nambu (Japan), corresponding to low, medium and high Peak Ground Acceleration (PGA), respectively. The response of nonlinear dynamic soil-structure system may be strongly affected by the time-domain character (e.g., frequency content, shape, number of pulses of time-history, and response spectrum characteristic) of time-histories even if the spectra of different time-histories are nearly identical.

## 6. Results and Discussion

Dynamic analyses were performed using the acceleration time-histories described above. The results obtained from DIANA were compared with those determined using a pseudo-static method (i.e. following the approach by Mononobe-Okabe). Following Green and Ebeling [11] approach, the dynamically-induced lateral earth pressures acting on the wall were computed by assuming constant stresses within the element. The corresponding lateral earth pressure coefficient ($K_{j,DIANA}$) could then be back-calculated at time increment $j$ from DIANA results using the following expression:

$$K_{j,DIANA} = \frac{2 \cdot P_{j,DIANA}}{\gamma \cdot H^2 \cdot (1 - k_{v,j})} \tag{5}$$

where, $P_{j,DIANA}$ is the resultant of force computed by DIANA and acting on the wall, $\gamma_t$ is the total unit weight of the backfill, $H$ is the height of the wall, and $k_{v,j}$ is the vertical inertial coefficient (assumed in this study equal to zero). Equation (5) is used to compute $K_{DIANA}$ values at times corresponding to the peaks in the time-history of the horizontal inertial coefficient ($k_h$), which is calculated from acceleration time-history recorded at the midpoint of the sliding wedge in both the active and passive sides. The direction of $k_h$ is opposite to direction of the acceleration and acts towards and away from the backfill (more detail can be found in Rajeev [12]). A plot of the computed $K_{j,DIANA}$ values versus $k_h$ is shown in Figure 4. The calculation of $K_{j,DIANA}$ was carried out at the preselected times, where the maximum acceleration occurs in all three time-histories to cover the entire range of $k_h$. Also shown in this figure are the lateral dynamic earth pressure coefficients (active: $K_{AE}$; Passive: $K_{PE}$) computed using the Mononobe-Okabe expressions for the wall-soil system (Okabe [5]; Mononobe [1]) and Wood [12] solution for rigid wall.

Following observations were made from the Figure 4.

1. Active pressure coefficient:
    a. $K_{DIANA} \approx K_{Mononobe-Okabe}$, for moderate levels of shaking
    b. $K_{DIANA} < K_{Mononobe-Okabe}$, for larger levels of shaking
    c. $K_{away\ from\ backfill} > K_{towards\ backfill}$
    d. The computed $K$ values show a general scatter around the curve for the Mononobe-Okabe dynamic active earth pressure curve.
2. Passive pressure coefficient:
    e. The computed $K_{DIANA}$ values for lower levels of shaking show values significantly lower than the $K_{Mononobe-Okabe}$
    f. The computed $K_{DIANA}$ increases with level of shaking
    g. The computed $K_{DIANA}$ values do not show a general scatter around the curve for the Mononobe-Okabe dynamic passive earth pressure curve

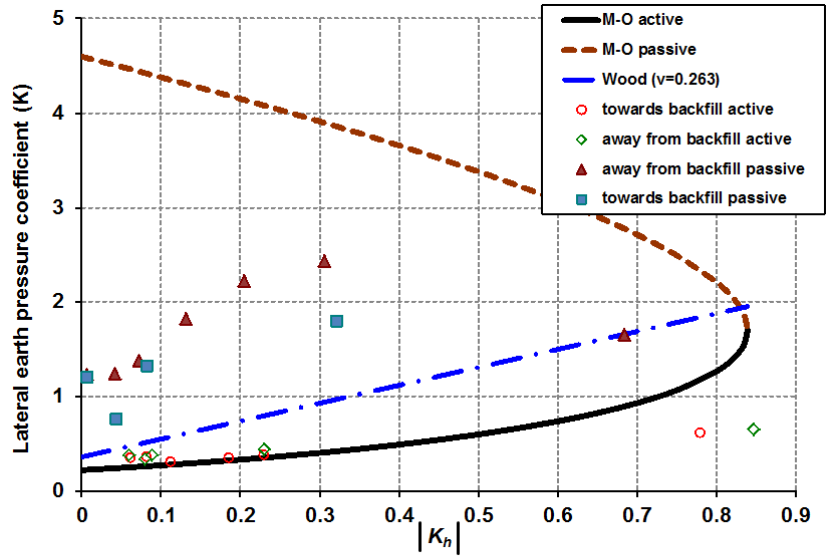

**Figure 4. Comparison of active and passive lateral earth pressure coefficients ($K_{DIANA}$) back-calculated from DIANA results with values computed using the Mononobe-Okabe and Wood expressions**

At larger levels of shaking, the Mononobe-Okabe expressions for active pressures failed to predict the induced stresses on the wall. The computed dynamic stresses from numerical analysis are higher than those computed by the Mononobe-Okabe equation for active pressures in the range of small to moderate levels of shaking.

At the lower levels of shaking, the passive pressures are not fully mobilized, therefore the $K$ values computed from the DIANA results are smaller than those calculated from the Mononobe-Okabe expressions. When the level of shaking increases, the mobilization of passive pressure also increases, and consequently the $K$ values increase.

The points of application of the total and incremental dynamic resultant forces are also important parameters for design and stability assessment of sheet pile wall. Therefore, the vertical distances ($Y_j$) from the base of the retaining wall to the points of application of total dynamic forces ($P_j$) acting on the wall were computed using the following relation:

$$Y_j = \frac{\sum_i h_j \cdot \sigma_{i,j} \cdot y_i}{\sum_i h_j \cdot \sigma_{i,j}} \quad (6)$$

where, $Y_j$ is the vertical distance from the base of the retaining wall to the point of application of the total resultant force acting on the wall at time increment $j$, $y_i$ is the vertical distance from the base of the retaining wall to the centre of element $i$, $\sigma_{i,j}$ is the average stress acting on the element $i$ and at time increment $j$, and $h_i$ is the length of element $i$

Figure 5 compares the point of application of total dynamic force together with point of application of static force calculated at the end of the construction of the wall. The point of application of static force (0.295%) calculated was below the value (0.33%) calculated using triangular stress distribution along the height, because the stresses below the point of rotation of wall had very large passive stresses. Further, the points of application of dynamic forces for lower levels of shaking show a scatter around 0.25% (H/4), but for larger level of shaking it showed big range of deviation.

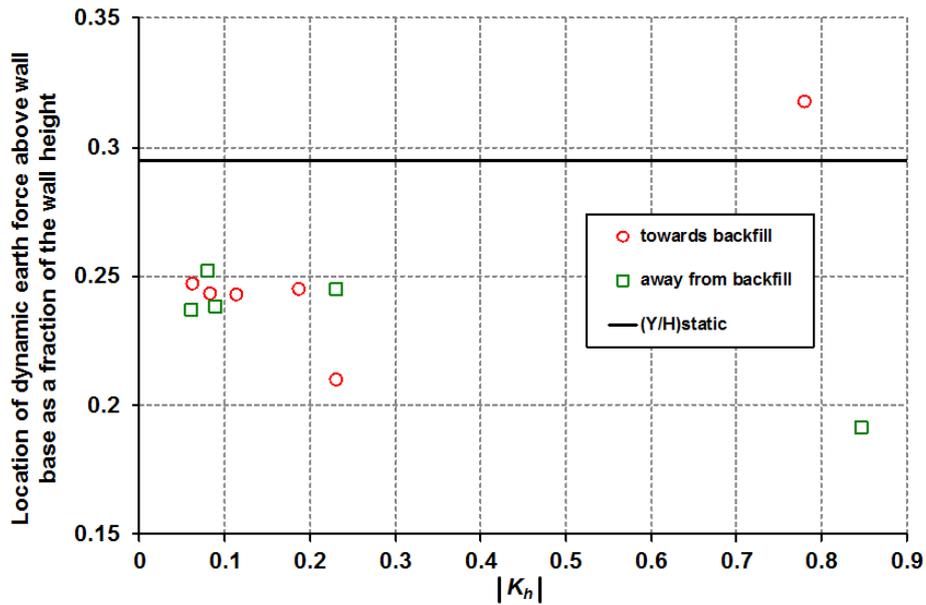

Figure 5. Point of application of total active dynamic force on sheet pile wall

## 7. Summary and Conclusions

This paper illustrates a preliminary investigation onto the seismic behaviour of flexible sheet pile wall, retaining a dry granular backfill. A finite element model of the system is built using DIANA finite element program. For the soil surrounding the sheet pile was modelled as elasto-plastic material with Mohr-Coulomb failure criterion.

The results from this study show that using the Mononobe-Okabe equation to design the sheet pile wall will provide good estimate for dynamic stresses on the active side. Conversely, it will overestimate dynamic stresses in the passive side. Further, the point of application of total dynamic forces in active side is below the point of application of static forces, around $0.25H$, because of the large passive pressure beyond the point of rotation of the wall.

The conclusions drawn from this study may not apply to retaining wall system of differing geometry and/or material properties. Further research is required to draw more general conclusions regarding the appropriateness of the Mononobe-Okabe method to evaluate the dynamic pressure induced under seismic conditions on the sheet pipe walls.